\documentclass[twocolumn]
{aastex631}
\usepackage{amsmath}
\usepackage{amssymb}
\usepackage{aas_macros}

\begin{document}

\title{Contribution of the Cygnus bubble to the Galactic cosmic ray spectrum and diffuse $\gamma$-ray emissions}

\correspondingauthor{Xiang-Li Qian}
\email{qianxl@sdmu.edu.cn}

\author{Lin Nie}
\affiliation{School of Mechanics and Aerospace Engineering, Southwest Jiaotong University, Chengdu, 610031, China}
\affiliation{School of Physical Science and Technology, Southwest Jiaotong University, Chengdu, 610031, China}
\affiliation{Department of Astronomy, Yunnan University, and Key Laboratory of Astroparticle Physics of Yunnan Province, Kunming, 650091, People’s Republic of China
}
\affiliation{Key Laboratory of Particle Astrophysics, Institute of High Energy Physics, Chinese Academy of Sciences, Beijing, 100049, China
}

\author{Xiang-Li Qian}
\affiliation{School of Intelligent Engineering, Shandong Management University, Jinan, 250357, China}


\author{Yi-Qing Guo}
\affiliation{Key Laboratory of Particle Astrophysics, Institute of High Energy Physics, Chinese Academy of Sciences, Beijing, 100049, China
}

\author{Si-Ming Liu}
\affiliation{School of Physical Science and Technology, Southwest Jiaotong University, Chengdu, 610031, China}

\begin{abstract}
Since the discovery of cosmic rays (CRs) over a century ago, their origin has remained a mystery and a key research question. Recently, the LHAASO experiment identified the first CR super-acceleration source, the Cygnus bubble, which can accelerate CRs to energies exceeding $\rm 10~PeV$. A pertinent question is: how much does the Cygnus bubble contribute to the CR spectrum observed on Earth? With the aim of answering that question, a 3D propagation analysis was conducted on CRs in this study. The Cygnus bubble was incorporated into our propagation model in order to determine its contributions to the observed spectra. First, we calculated the spectrum and spatial morphology of the Cygnus bubble to reproduce the observed LHAASO data. Subsequently, we calculated the diffuse $\gamma$-ray emissions produced by the CRs from the Cygnus bubble and the energy spectrum of the cosmic ray particles near Earth after propagation. Finally, we utilized a CR spatial-dependent propagation model to calculate the large-scale CR energy spectrum and the resulting diffuse $\gamma$-ray emissions. Our results indicate that: (1) the Cygnus bubble contributes minimally to the CR spectrum observed on Earth, (2) the emissions produced by the CR particles from the Cygnus bubble dominates the diffuse $\gamma$-ray emissions in that region, (3) the structural fluctuations of the diffuse $\gamma$-ray emissions observed by LHAASO are likely due to the local CR halo. We anticipate that LHAASO will identify more CR halo sources to validate our model.

\end{abstract}


\section{Introduction} \label{sec:intro}
The origin of cosmic rays (CRs), which are high-energy particles from outer space, has been a fundamental question in 21st century physics that holds significant scientific importance. Since their discovery in 1912, CRs have raised unresolved questions regarding their origin, acceleration mechanisms, and propagation processes. The widely accepted view is that CRs with energies below the ``knee" region ($\rm 3\times10^{15}~eV$) originate from Galactic sources. Possible sources of Galactic CRs include supernova remnants\citep{2013Sci...339..807A,2014IJMPD..2330013A,2023ApJ...952..100N}, pulsar wind nebulae (as part of supernova remnants)\citep{2021Natur.594...33C,2021Sci...373..425L,2022ApJ...924...42N}, binary star systems\citep{2012Sci...335..175M}, superbubbles\citep{2024SciBu..69..449L}, and the Galactic center\citep{2016Natur.531..476H}.  Among these possibilities, supernova remnants are the most likely sources\citep{2013A&ARv..21...70B}. These remnants are considered the primary sources of Galactic CRs in standard propagation models, whereby CRs are primarily accelerated through nonlinear diffusive shock acceleration mechanisms\citep{2002APh....16..429B,2004APh....21...45B,2012APh....39...12Z,2010ApJ...718...31P}. Once accelerated, the CR particles escape from their sources, diffuse into the interstellar medium (ISM), and eventually permeate the entire Milky Way\citep{2007ARNPS..57..285S,2013PhPl...20e5501Z}. However, since most CR particles are charged high-energy nucleons, they are deflected by the Galactic or local turbulent magnetic fields during propagation. Consequently, the CR particles lose their directional information and exhibit an almost isotropic distribution. Therefore, it is challenging to pinpoint the locations of the CR sources solely by observing the CR particle spectrum. Nonetheless, there may be another way to identify the sources of CRs.  CRs interact with the ISM through proton-proton (pp) interactions during acceleration and propagation. These interactions produce neutral emissions, namely electromagnetic radiation and neutrino emissions. These neutral emissions are not influenced by magnetic fields and propagate in straight lines, making them valuable tools for identifying CR sources.

Investigating the mechanism of $\gamma$-ray emissions from CR sources can offer direct evidence for the origins of CRs. However, distinguishing whether the observed high-energy $\gamma$-rays originate from leptonic or hadronic processes remains a significant challenge. Researchers anticipate that observing higher energy thresholds will provide direct evidence of hadronic sources, thus unraveling the century-old mystery of where CRs originate from.

Although researchers have endeavored to identify hadronic sources through experimental observations, to date, only a few sources can be adequately explained using hadronic models. For example, in 2013, the Fermi Gamma-ray Space Telescope detected a characteristic ``bump” around $\rm 70~MeV$ in the spectra of supernova remnants IC443 and W44, which are indicative of $\pi^0$ meson decay\citep{2013Sci...339..807A,2011ApJ...742L..30G}. H.E.S.S. observed high-energy gamma photons from the Galactic center with energies above $\rm 10~TeV$, and the emissions correlate with the distributions of molecular clouds in the vicinity. To further validate the emission mechanisms of these sources and identify CR sources, higher-energy $\gamma$-ray observations are urgently needed.

LHAASO is one of the most significant scientific facilities for the study of CRs, since it provides the highest sensitivity for observing $\gamma$-rays exceeding $\rm 20~TeV$. Recently, LHAASO reported electromagnetic emissions from the Cygnus bubble region for the first time, with energies up to a few PeV\citep{2024SciBu..69..449L}. These emissions correlate with the gas distribution in this region, suggesting a hadronic origin. The central region of the Cygnus bubble has a relatively concentrated gamma photon distribution, with a significantly higher photon density than in other regions of the bubble. This indicates the presence of a CR accelerator within the bubble that continuously injects CRs into the surrounding region. Additionally, LHAASO’s latest observations of the diffuse $\gamma$-ray emissions in the Galaxy disk revealed a ``bump” structure in the vicinity of the Cygnus bubble (Galactic longtitude 60--80 degrees)\citep{lhaaso2024ultrahigh,2024PhRvD.109f3001Y}, which contrasts sharply with the ``dip" structure predicted by both the standard and CR spatial-dependent propagation (SDP) models.

Based on these latest observations, questions naturally arose regarding whether the “bump” structure observed by LHAASO originated from the Cygnus bubble and also whether the CRs accelerated in the Cygnus bubble contributed to the CRs observed near Earth. This study primarily uses the Cygnus bubble $\gamma$-rays observed by LHAASO to constrain the injection of CR particles. The aim of this study is to investigate whether the emissions from CR particles escaping from the Cygnus bubble contributes to the local diffuse $\gamma$-ray emissions in the Galaxy observed near Earth, and whether the Cygnus bubble contributes to the CR proton spectrum observed on Earth. This paper is structured as follows: Section \ref{sec:method} outlines the research methodology and the CR SDP model; Section \ref{sec:result} presents and discusses our results; and Section \ref{sec:conclusion} provides a summary and discusses future prospects of this study.

\section{Methodology and Model} \label{sec:method}
We propose that the Galactic diffuse $\gamma$-ray emissions observed by LHAASO are a result of combined contributions from background CRs and local sources. Therefore, we employed a SDP model in this study to simulate the distribution of Galactic background CRs and the resultant diffuse $\gamma$-ray emissions. Subsequently, we calculated the emission spectrum produced by the CRs within the Cygnus bubble to interpret the latest observational data from LHAASO. Finally, we excluded the emissions from point and extended sources within the Cygnus bubble and investigated the contribution of Cygnus bubble $\gamma$-ray emissions to the overall diffuse $\gamma$-ray emissions.

\subsection{Propagation Model of Background CRs} \label{sec2.1}
The observed Galactic CRs are thought to arise from a combination of two fundamental astrophysical plasma phenomena: the acceleration of CR particles and their transport through a turbulent medium. The interaction of these acceleration and diffusion processes results in a primary CR spectrum described by $\phi_{\mathrm{p}}\sim\mathrm{Q} / \mathrm{K}\propto\mathrm{R}^{-\nu-\delta}$ and a secondary-to-primary ratio spectrum described by $\phi_{\mathrm{s}} / \phi_{\mathrm{p}} \sim \mathrm{R}^{-\delta}$\citep{2015PhRvD..92h1301T}. Traditional CR propagation models assume that CR diffusion is both uniform and isotropic.

While traditional CR propagation models assume that CR diffusion is both uniform and isotropic, our proposed spatial-dependent model(SDP) differs in certain aspects. We propose that CR sources are primarily distributed within the Galactic disk. In the vicinity of these sources, there often exists a slow diffusion zone, known as the halo region. These sources within the Galactic disk act as barriers, preventing external CRs from penetrating. They also act as potential wells, restricting the escape of local CRs. Thus, the spatial dependence of CR diffusion within the Galaxy is included in our model. The local CR spectrum is primarily composed of two components\citep{2024PhRvD.109f3001Y}.  One component is from distant CR sources that accelerate and propagate.  Due to their dispersion through the outer disk, these emissions exhibit a soft spectrum that primarily contributes to the low-energy component, known as the ``CR sea”. The other component originates from local sources and possesses a harder spectrum that dominates the high-energy component. The propagation process of CRs is described by the following diffusion equation:\citep{2007ARNPS..57..285S,2017JCAP...02..015E}:
\begin{equation}
    \begin{aligned}
        \frac{\partial \psi(\vec{r}, p, t)}{\partial t}= & Q(\vec{r}, p, t)+\vec{\nabla} \cdot\left(D_{x x} \vec{\nabla} \psi-\vec{V}_c \psi\right) \\
        & +\frac{\partial}{\partial p}\left[p^2 D_{p p} \frac{\partial}{\partial p} \frac{\psi}{p^2}\right] \\
        & -\frac{\partial}{\partial p}\left[\dot{p} \psi-\frac{p}{3}\left(\vec{\nabla} \cdot \vec{V}_c\right) \psi\right]-\frac{\psi}{\tau_f}-\frac{\psi}{\tau_r}
        \end{aligned}
\end{equation}
where $\psi(\vec{r}, p, t)$ represents the CR density per unit of total particle momentum $p$ at position $\vec{r}$, $Q(\vec{r}, p, t)$ describes the source term, $D_{x x}$ denotes the spatial diffusion coefficient, $\vec{V}_c$ is the convection velocity, and $\tau_f$ and $\tau_r$ are the timescales for loss by fragmentation and radioactive decay, respectively. In the outer disk, CR diffusion is described by traditional models, i.e., $D_{x x}(\mathcal{R})=D_0\left(\frac{\mathcal{R}}{\mathcal{R}_0}\right)^{\delta_0}$. Within the Galactic disk, however, CR diffusion depends on the distribution of CR sources $f(r,z)$, and the diffusion coefficient is described as\citep{2016ApJ...819...54G,2018PhRvD..97f3008G} 

\begin{equation}
D_{x x}(r, z, \mathcal{R})=D_0 F(r, z) \beta^\eta\left(\frac{\mathcal{R}}{\mathcal{R}_0}\right)^{\delta_0 F(r, z)},
\end{equation}
where $\delta_0 F(r, z)$ describes the turbulent characteristics of the local medium environment and $D_0 F(r, z)$ represents the normalization factor of the diffusion coefficient at the reference rigidity.

\begin{equation}
    F(r, z)= \begin{cases}g(r, z)+[1-g(r, z)]\left(\frac{z}{\xi z_0}\right)^n, & |z| \leq \xi z_0 \\ 1, & |z|>\xi z_0\end{cases},
\end{equation}
here, $\xi z_0$ denotes the half-thickness of the Galactic halo, and $g(r, z)=N_m /[1+f(r, z)]$, where $N_m$ is the normalization factor. The parameter $n$ is used to describe the smoothness between the inner and outer halos, and the source distribution $f(r,z)$ is a cylindrically symmetric continuous distribution\citep{1996A&AS..120C.437C,1998ApJ...509..212S,1998ApJ...504..761C}, 
\begin{equation}
f(r, z)=\left(\frac{r}{r_{\odot}}\right)^{1.25} \exp \left[-\frac{3.87\left(r-r_{\odot}\right)}{r_{\odot}}\right] \exp \left(-\frac{|z|}{z_s}\right),
\end{equation}
Where $r_{\odot}$ = 8.5 kpc and $z_s$ = 0.2 kpc. 

We assume that the CR injection spectrum follows a rigidity cutoff power-law distribution. Utilizing the GALPROP numerical package\citep{2022ApJS..262...30P,1998ApJ...509..212S}, we solved the CR propagation equations and calculated the diffuse $\gamma$-ray emissions produced by the CRs.

\subsection{Diffusion in the Cygnus Bubble} \label{sec2.2}
The Cygnus bubble region is considered a slow diffusion zone. Within this bubble, a CR source resembles a point source that accelerates and continuously emits CRs. Due to the smaller diffusion coefficient of the CR particles in the bubble, a potential well is formed and it limits the rapid diffusion of CR particles. Thus, we modelled the diffusion of CR particles in the Cygnus bubble as a ``two-zone" diffusion, consisting of slow diffusion within the bubble and relatively faster diffusion in the ISM. Mathematically, the diffusion coefficient is described as follows\citep{2019ApJ...879...91J}:
\begin{equation}
D=\beta\left(\frac{\mathcal{R}}{\mathcal{R}_0}\right)^{\delta} \begin{cases}D_b, & r<r_i \\ D_b\left[\frac{D_0}{D_b}\right]^{\frac{r-r_i}{r_o-r_i}}, & r_i \leqslant r \leqslant r_o \\ D_0, & r>r_o\end{cases}
\end{equation}
Where, $\beta$ represents the particle velocity in units of the speed of light, $D_0$ is the normalization of the diffusion coefficient in the general ISM, $D_b$ is the normalization for the diffusion coefficient within the SDZ with radius $r_i$, $R$ is the particle rigidity, and $R_0$ is the normalization (reference) rigidity. The zone between $r_i$ and $r_o$ is a transition layer where the normalization of the diffusion coefficient increases exponentially from $D_b$ to the ISM value $D_0$.

Following the scenario proposed in the LHAASO paper\citep{lhaaso2024ultrahigh}, we assumed that the CR protons are injected into the bubble by a point source at a constant rate. The injection spectrum follows a power-law distribution with a high-energy cutoff, $Q \propto  Q_0 E_{\mathrm{p}}^{-s} \exp \left(-E_{\mathrm{p}} / E_0\right)$, where $E_0$ represents the cutoff energy of the spectrum and $Q_0$ is the normalization of the injected protons.

\section{Results} \label{sec:result}
In this section, we present our computational results. First, we reproduced the LHAASO observations of the Cygnus bubble to constrain the escaped CR protons within the bubble. Then we calculated the diffuse $\gamma$-ray emissions produced by the CR protons originating from the bubble. Finally, we evaluated the contribution from the Cygnus bubble CR protons that diffused into the vicinity of Earth after being injected into the ISM.

\subsection{Emissions from the Cygnus bubble}
\begin{figure}[t]
\centering
\includegraphics[width=0.95\linewidth]{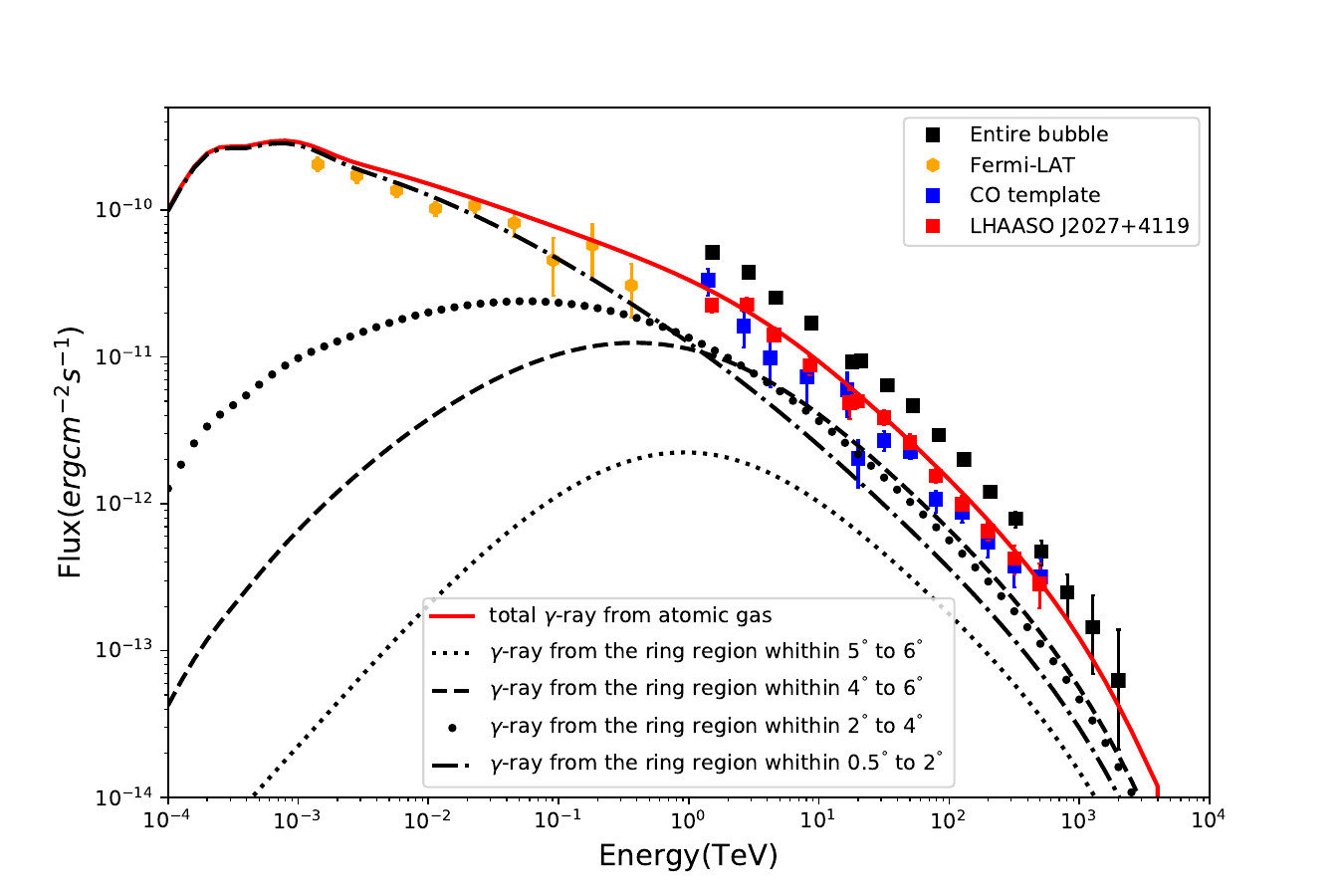}
\caption{illustrates the emissions from the Cygnus bubble. The red line represents the multi-wavelength non-thermal radiation spectrum across the entire bubble region accompanied by the corresponding observation. The black lines represent the contributions of the $\gamma$-rays from different ring regions. The orange points denote the Fermi-LAT observations\citep{2019NatAs...3..561A} from the Cygnus Cocoon. The black, red, and blue points represent the flux detected by LHAASO\citep{2024SciBu..69..449L} from the entire bubble, LHAASO J2027+4119, and the CO gas template, respectively.}
\label{fig1}
\end{figure}

\begin{figure*}[htbp]
\centering
\includegraphics[width=0.3\linewidth]{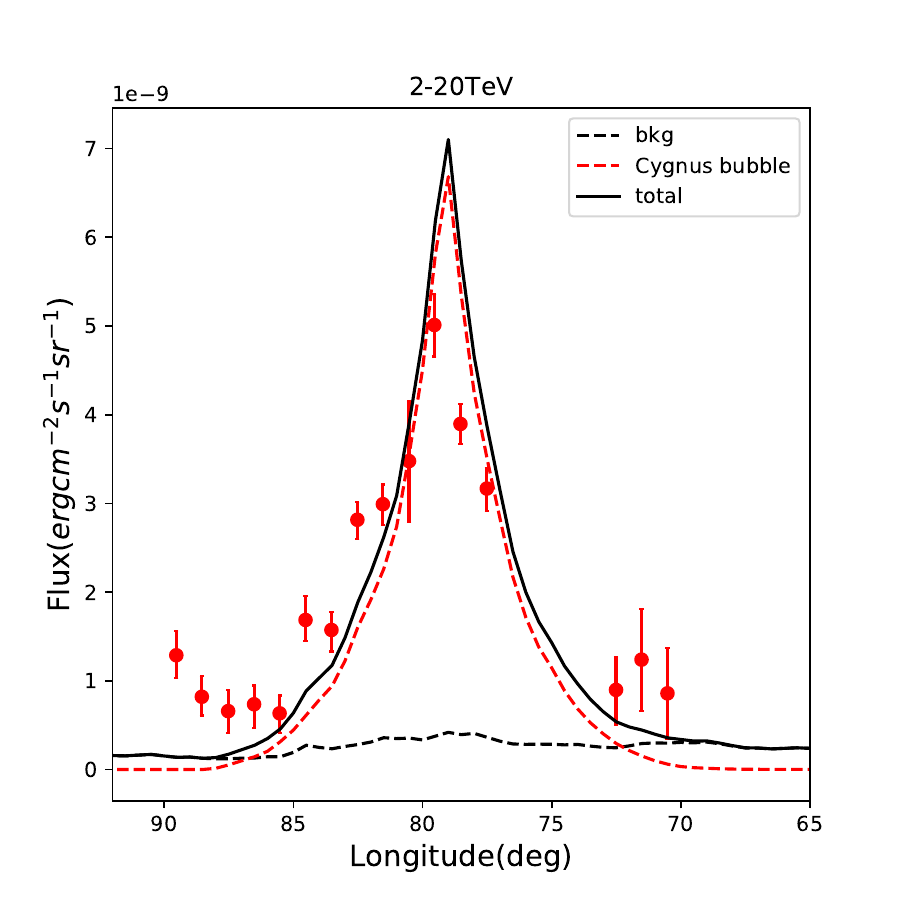}
\includegraphics[width=0.3\linewidth]{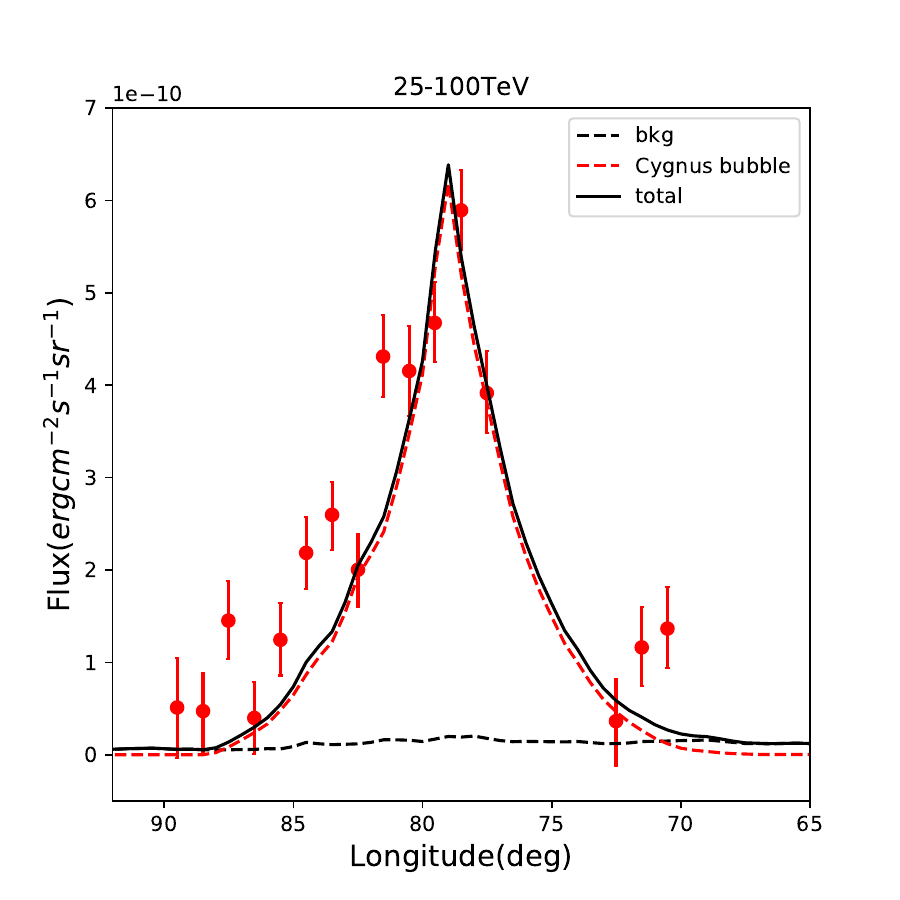}
\includegraphics[width=0.3\linewidth]{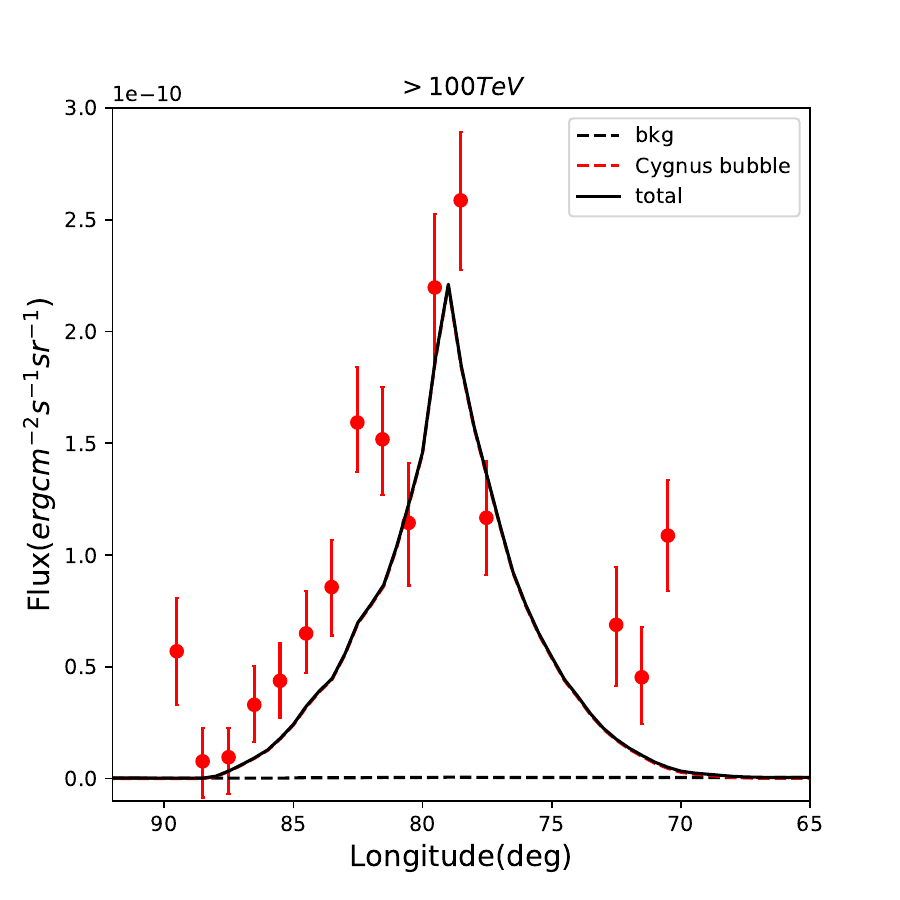}
\caption{illustrates the distribution of gamma-ray emissions from the Cygnus bubble combined with observations from LHAASO\citep{2024SciBu..69..449L}. Each subplot shows the distribution of gamma-ray emissions with Galactic longitude in different energy bands within the latitude range of -2° to 2°. The red dashed lines indicate the components from CR protons produced by the Cygnus bubble and the black dashed lines show the background diffuse $\gamma$-ray components, respectively.}
\label{fig2}
\end{figure*}

To calculate the contribution of diffuse $\gamma$-rays generated by the CRs from the Cygnus bubble and the contribution of CR protons near Earth after propagation, we employed the methods described in Section \ref{sec2.2} to compute the emission spectrum and its morphology from the bubble. As shown in Figure \ref{fig1} and \ref{fig2}, the LHAASO and Fermi-LAT observations can be well reproduced. Figure\ref{fig1} shows the emissions produced by the cosmic ray protons  within the different extended regions around the bubble core through  p-p interaction process. While Figure\ref{fig2} describes the distribution of gamma ray emissions as function of the galactic longitude with a latitude from -2° to 2°. Figure\ref{fig1} and \ref{fig2} describes the radiation characteristics of the Cygnus bubble.
For the injection spectrum index of the acceleration sources in the bubble and the spectral index of CR particle diffusion, we adopted the values fitted by the paper \citep{2024SciBu..69..449L}, i.e., $s$ = -2.25 and $\delta$ = 0.7. It should be noted that the size of the bubble affects the spectral shape of the spectral energy distribution produced by the CR protons within the bubble. The size of the bubble also determines the radial scale of the $\gamma$-ray distribution of the bubble, particularly in the high-energy range. This is because high-energy CR particles have a longer diffusion distance for the same evolutionary time range. Consequently, reducing the size of the bubble decreases the number density of high-energy CR particles, thereby softening the overall radiative spectrum of the bubble at high energies. This is consistent with observations showing that the spectral data softened above approximately $\rm 1~TeV$. The ratio of the diffusion coefficient within the bubble to that in the ISM, along with the bubble size, jointly influences the emission spectrum. Therefore, by fitting the observed spectrum and morphology of the bubble's emissions, 
we obtained $D_b=4.0\times10^{25}cm^2s^{-1}$. But the reference rigidity was set to $\rm R_0=1~TV$ in the bubble, consistent with the value used in the LHAASO paper\citep{2024SciBu..69..449L}. The size of the extended bubble is about 10°, corresponding to a physical distance of approximately 400 parsecs.

\subsection{Diffuse gamma-ray emissions from the Cygnus bubble and large-scale CRs}
After reproducing LHAASO’s observations of the Cygnus bubble, we calculated the diffuse $\gamma$-ray emissions produced by the CR protons from this bubble. The CR proton energy spectrum after propagation to Earth's vicinity was also calculated. To replicate the diffuse $\gamma$-ray spectrum observed by LHAASO, it was necessary to determine the distribution of background CRs and their resultant diffuse $\gamma$-ray emissions. Using the methodology described in Section \ref{sec2.1}, we simulated the propagation of background CRs. The diffusion coefficient in our model was refined to better align with experimental observations. It was then possible to calculate the density of the CR protons and their diffuse emissions at various locations within the Galaxy.

\begin{figure}[t]
\centering
\includegraphics[width=0.95\linewidth]{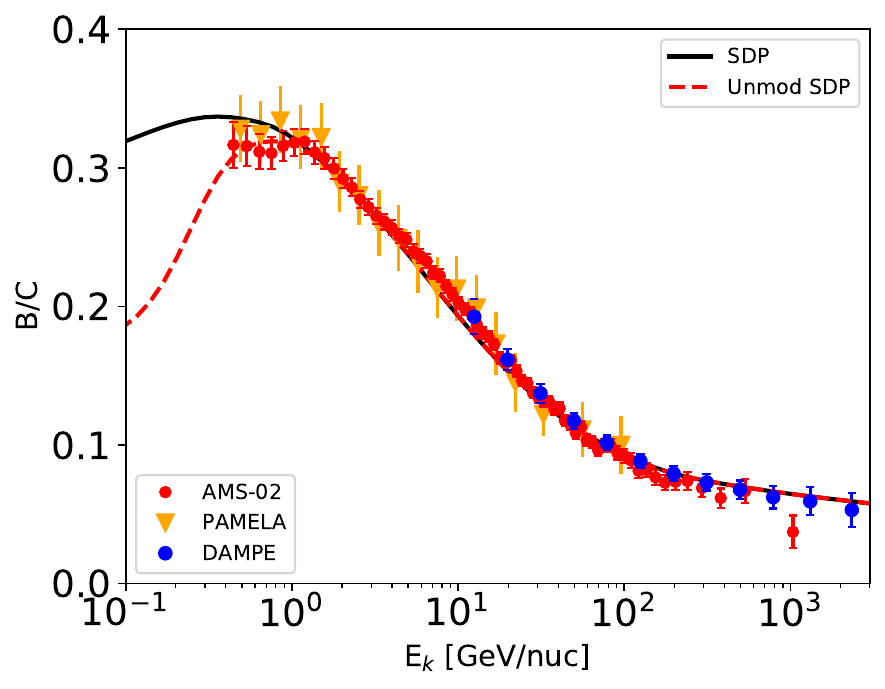}
\caption{compares the B/C ratio calculated using the CR SDP model with the observational data from AMS-02\citep{2017PhRvL.119y1101A}, PAMELA\citep{2014ApJ...791...93A}, and DAMPE\citep{2022SciBu..67.2162D}. The red dashed line represents the spectrum without accounting for solar modulation. In this paper, the solar modulation potential is uniformly considered as 550 MeV.}
\label{fig3}
\end{figure}

\begin{table}[b]
\footnotesize
\centering
\caption{Parameters of the SDP model.}
\label{tab1}
\tabcolsep 3.5pt
\begin{tabular*}{0.47\textwidth}{lcccccc}    
\hline \hline
$D_0{ }\left[\mathrm{cm}^{2} \mathrm{~s}^{-1}\right]$ & $\delta_0$ & $N_m$ & $\xi$ & $\mathrm{n}$ & $v_A\left[\mathrm{~km} \mathrm{~s}^{-1}\right]$ & $z_0[\mathrm{kpc}]$ \\
\hline $4.0 \times 10^{28}$ & 0.63 & 0.24 & 0.1 & 4.0 & 6 & 4.5 \\
\hline 
\end{tabular*}
\end{table}

\begin{figure}[t]
    \centering
    \includegraphics[width=0.95\linewidth]{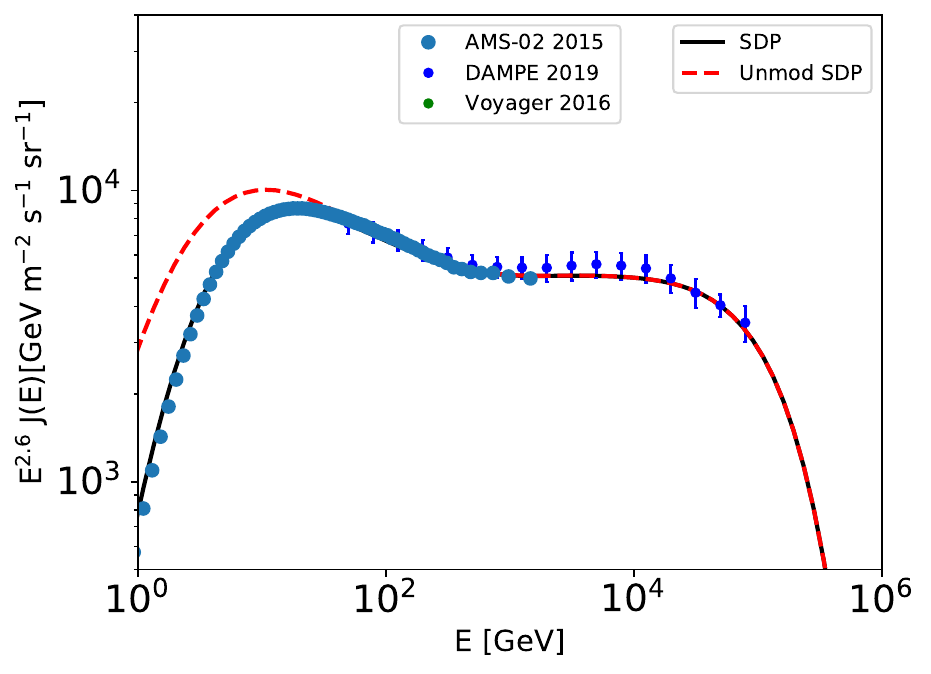}
    \caption{shows the CR proton spectrum calculated by the CR SDP model and the observational data from AMS-02\citep{2015PhRvL.114q1103A} and DAMPE\citep{2019SciA....5.3793A}. The red dashed line represents the spectrum without accounting for solar modulation.}
    \label{fig4}
    \end{figure}

The diffusion coefficient parameters were determined using the B/C flux ratio of the observational data. By fixing the halo height H at 4.5 kpc, a value of $\delta_0=0.63$ was obtained. The Alfv$\acute{e}$n wave and other relevant model parameters were determined by fitting the low-energy B/C flux ratio and CR proton observations, with these parameters listed in Table \ref{tab1}. Figures \ref{fig3} and \ref{fig4} show the theoretical and observational data from Earth of the CR B/C flux ratio and proton spectra, respectively. As can be seen, the CR SDP model accurately reproduces the observed CR B/C flux ratio and proton spectra. The diffuse $\gamma$-ray emissions produced by the background CR particles were also calculated concurrently. In line with LHAASO’s method for procession observational data, contributions from point sources and extended sources were subtracted to calculate the diffuse $\gamma$-ray emissions\citep{2024SciBu..69..449L}, when the components of Cygnus bubble and background were calculated, respectively. Consequently, the sources detected by KM2A and others aggregated in TeVCat were masked during the calculation. Specifically, the focus of this study, the Cygnus bubble, was masked over a region of approximately 6°, which is approximately 2.5 times larger than other sources in the TeVCat database. 

\begin{figure}[t]
    \centering
    \includegraphics[width=0.95\linewidth]{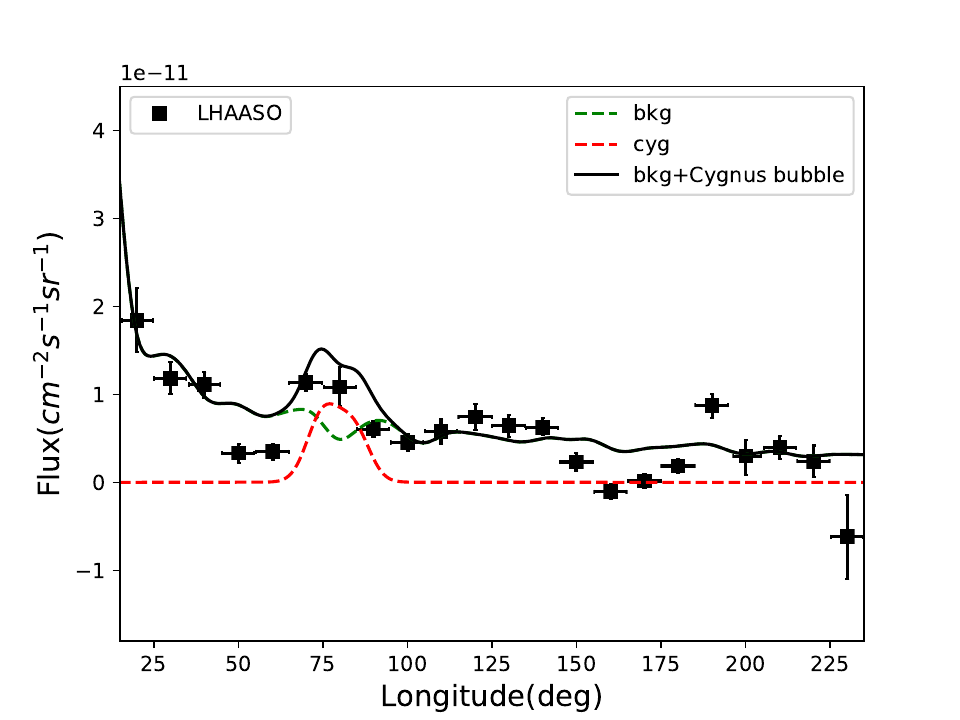}
    \caption{shows the flux of diffuse gamma-ray emissions in the Galactic longitude range of $\lvert$b$\rvert$$<$5° with energy range 10-63 TeV, in comparison with the LHAASO observations\citep{lhaaso2024ultrahigh}. The red dashed line represents the diffuse gamma components from the CR protons produced by the Cygnus bubble, and the green dashed line represents the emission flux of diffuse gamma predicted by the CR SDP model.}
    \label{fig5}
    \end{figure}

Figure \ref{fig5} presents the flux of the model-predicted diffuse $\gamma$-ray emissions, the component produced by the CR protons from the Cygnus bubble, and the observational data from LHAASO. It can be observed that around 60°-80° Galactic longitude, the model-predicted flux of diffuse $\gamma$-ray emissions underestimates the LHAASO’s observations, with the spectral trend moving inversely. However, the component contributed by the CR protons from the Cygnus bubble closely matches the LHAASO observations in this region, indicating that the Cygnus bubble significantly contributes to the surrounding diffuse emissions. Therefore, it can be inferred that the notable underestimations seen in the model's predictions compared to LHAASO’s data in other regions may be due to a local CR source dominating the diffuse emissions in those regions. For instance, the observational data exhibit a bump around the 190°-200° region, which can possibly be attributed to the incomplete subtraction of emissions produced by the local Geminga halo. However, detailed calculations in that domian are beyond the scope of this paper and will be discussed in future works.

\begin{figure}[t]
    \centering
    \includegraphics[width=0.95\linewidth]{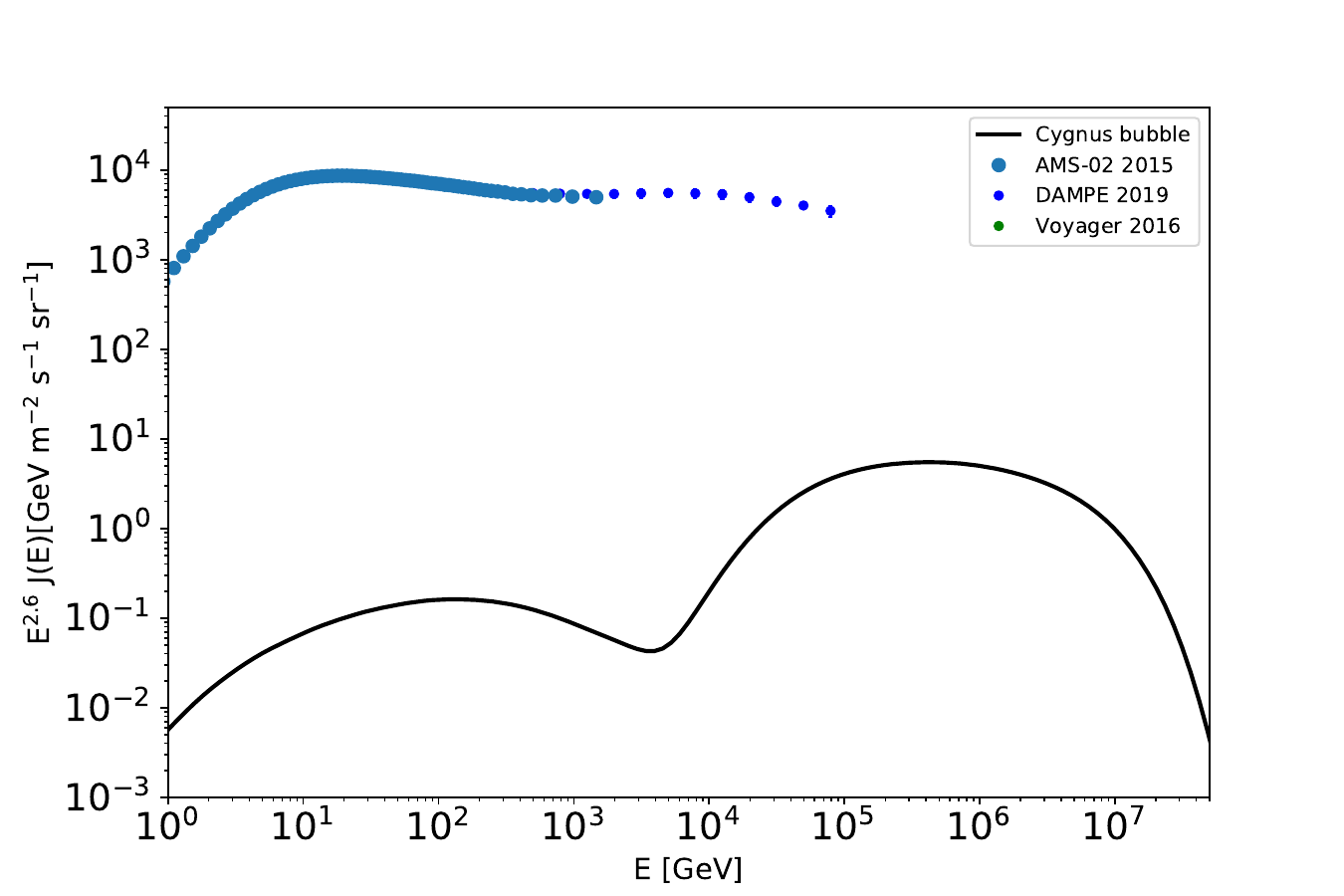}
    \caption{presents the CR proton flux from the Cygnus bubble as expected on Earth in comparison with the observed data.}
    \label{fig6}
    \end{figure}
\subsection{Contribution of the Cygnus bubble to the CR proton spectrum}
The energy spectrum of the CR protons from the Cygnus bubble near Earth was calculated based on the constraint from LHAASO’s extensive observational data, including the emission spectrum, radial distribution of the Cygnus bubble, and the diffuse $\gamma$-rays surrounding it. As illustrated in Figure \ref{fig6}, the Cygnus bubble's contribution is at least two orders of magnitude smaller than the observed data. At approximately 10 TeV, the number density of protons starts to increase significantly. This is because the bubble is a slow-diffusion region where low-energy cosmic ray particles are confined for longer periods and lose energy through radiative processes. This is consistent with the radiation within the bubble. Increasing the size of the bubble or the diffusion within the bubble will harden the radiation spectrum at high energies and cause a situation where the proton spectrum's number density begins to increase and shift to higher energies. Notably, when calculating the propagation of CR protons from the Cygnus bubble in the ISM, we used the traditional CR propagation model with the spectral index of the diffusion coefficient set to the classical Kolmogorov-type turbulence with a value of $\delta_0$ = 0.33. Our findings suggest that the Cygnus bubble has a negligible impact on the CR spectrum near Earth. However, this source significantly contributes to the diffuse $\gamma$-ray emissions in its vicinity. Thus, our calculations support the conclusion that local sources are the primary contributors of high-energy CR components, with the local CR halo both preventing the propagation of external CRs and restricting the escape of local CRs.

\section{Conclusion and Summary} \label{sec:conclusion}
Based on the LHAASO observations and the GALPROP simulations, we investigated the spatial dependence of diffuse $\gamma$-ray emissions on CRs. Through this process, the accuracy of the CR SDP model was tested. 

LHAASO’s latest results include the first detection of high-energy emissions from the Cygnus bubble. These emissions reached up to several PeV and correlated with the local gas distribution. However, the notable deviations in the Galactic diffuse $\gamma$-ray emission’s longitudinal distribution from the gas distribution raised several questions. What is the contribution of the Cygnus bubble to the CR protons near Earth? Are the characteristics of the local Galactic diffuse emissions related to this bubble? In this study, a more realistic representation of the Cygnus bubble was incorporated into a 3D CR propagation model. The model successfully reproduced LHAASO’s observations of the Cygnus bubble’s energy spectrum and spatial morphology. We then examined whether the emissions produced by the CR protons from the Cygnus bubble contributes to the Galactic diffuse $\gamma$-ray emissions. Our results indicate that the diffuse $\gamma$-ray emissions from the CR protons of the Cygnus bubble dominates the emissions in this region. Therefore, it is suggested that the spatial distribution characteristics of the diffuse $\gamma$-ray emissions observed by LHAASO, and similar ``bubble" structures in different regions, may be due to the inability to completely subtract the emissions from local sources.

Conversely, the CR protons from the Cygnus bubble contribute negligibly to the CR proton spectrum observed near Earth. These findings reveal that the Galactic diffuse $\gamma$-ray emissions exhibits spatial dependence, with the CR halo acting both as a barrier preventing the propagation of external CRs and as a well restricting the diffusion of local CRs.

This study shows that the CR halo significantly influences local diffuse $\gamma$-ray emission. It is recommended that future works incorporate all observed TeV halos to calculate local source contributions in order to add more samples to the framework. It is also anticipated that LHAASO will conduct precise observations of more CR halos, which will further validate the viewpoints presented in this study.

\section*{Acknowledgements}
We thank  Prof. He Hui-hai for presenting the idea and helpful discussion. This work is supported by the National Natural Science Foundation of China (12263005,12275279,12375103), and Foundations from Yunnan Province (202301AS070073).

\bibliographystyle{aasjournal}
\bibliography{ref_v2}{}
\end{document}